\documentclass{iau} 
\usepackage{siunitx}
\usepackage{graphicx}
\usepackage{epstopdf}
\usepackage{verbatim}

\title[Neutrino \& $\gamma$-ray Emission from the Core of NGC1275] 
{Neutrino and $\gamma$-ray Emission from the Core of NGC1275 by Magnetic Reconnection:
 GRMHD Simulations and Radiative Transfer/Particle Calculations}

\author[Rodr\'iguez-Ram\'irez et al.]   
{J.C. Rodr\'iguez-Ram\'irez$^1$, E. M. de Gouveia Dal Pino$^1$ \& R. Alves Batista$^1$
}

\affiliation{$^1$
Instituto de Astronomia,  Geof\'isica e Ci\^{e}ncias Atmosf\'ericas (IAG-USP),
Universidade de S\~{a}o Paulo.
Cidade Universitaria
R. do Mat\~ao, 1226
05508-090   S\~ao Paulo, SP
Brasil
\\ email: {\tt juan.rodriguez@iag.usp.br} 
}

\pubyear{2018}
\volume{342}  
\jname{IAUS 342: Perseus in Sicily: from black hole to cluster outskirts
}
\editors{Keiichi Asada, Elisabete de Gouveia Dal Pino,\\
Hiroshi Nagai, Rodrigo Nemmen, \& Marcello Giroletti, eds.}
\begin{document}

\maketitle

\begin{abstract}

Very high energy (VHE) emission has been detected from the radio galaxy NGC1275, establishing it as a potential cosmic-ray (CR) accelerator and a high energy neutrino source. We here study neutrino and $\gamma$-ray emission from the core of NGC1275 simulating the interactions of CRs assumed to be accelerated by magnetic reconnection, with the accreting plasma environment. To do this, we combine (i) numerical general relativistic (GR) magneto-hydrodynamics (MHD), (ii) Monte Carlo GR leptonic radiative transfer and, (iii) Monte Carlo interaction of CRs. A leptonic emission model that reproduces the SED in the [$10^3$-$10^{10.5}$] eV energy range is used as the background target for photo-pion interactions+electromagnetic cascading. CRs injected with the power-law index κ=1.3 produce an emission profile that matches the VHE tail of NGC1275. The associated neutrino flux, below the IceCube limits, peaks at $\sim$PeV energies. However, coming from a single source, this neutrino flux may be an over-estimation.

\end{abstract}
\firstsection 
\section{Introduction}

Very high energy (VHE) emission has been detected from NGC1275 (Aleksic et al. 2012), 
establishing it as potential cosmic-ray (CR) accelerator as well as a source of high energy 
neutrinos. The non-blazar nature of this radio galaxy suggests that the high energy emission
could be produced in the outflows as well as in the accreting core region.
The characteristics of potential neutrino emission from this source
could be crucial for determining possible regions of CR acceleration.

In this work, we investigate the neutrino and $\gamma$-ray emission signatures of 
CRs accelerated by magnetic reconnection in the core region of NGC1275. 
This energy release mechanism has been proven to occur in accreting magnetised plasmas 
(de Gouveia Dal Pino \& Lazarian, 2005,
Singh, de Gouveia Dal Pino \& Kadowaki, 2015,
Kadowaki, de Gouveia Dal Pino \& Singh 2015,
Kadowaki, de Gouveia Dal Pino \& Stone, 2018a, 
Kadowaki, de Gouveia Dal Pino \& Stone, 2018b)
and to accelerate particles via first order Fermi reconnection acceleration 
(Kowal, de Gouveia Dal Pino \& Lazarian, 2011,2012,
del Valle, de Gouveia Dal Pino \& Kowal, 2016;
see also de Gouveia Dal Pino et al. these procs).

We here probe neutrino and $\gamma$-ray emission 
by means of Monte Carlo simulations of the interactions of CRs with the 
magnetic and photon fields produced by the accreting flaw in the core region of NGC1275.
For this aim, we combine three numerical techniques: (i) Numerical general relativistic (GR)
magnetohydrodynamics (MHD) to model the magnetised accreting flow, (ii) Monte Carlo
GR leptonic radiation transfer to obtain the environment radiation field, 
and (iii) 
Monte Carlo propagation and interaction of CRs
assumed to be accelerated by magnetic reconnection,
to probe neutrino and hadronic $\gamma$-ray emission due to photo-pion processes and electromagnetic cascades.

In the next section we describe the numerical model for the environment in the core of NGC1275 and  
in Section 3, the CRs propagation stimulations from which we obtain the signatures of neutrino and hadronic
$\gamma$-ray emission. Finally, in Section 4 we discuss our results and state our conclusions.

\section{Background ADAF Model: Numerical GRMHD + Leptonic Radiative Transfer}

To study hadronic emission from the core region of  NGC1275, we consider a numerical
GRMHD, advection dominated accretion flaw (ADAF) model, as the environment where CRs (assumed to be accelerated by magnetic reconnection) and their secondary particles
propagate and interact.

The background magnetic field is obtained employing the axi-symmetric version of the publicly available
{\tt harm} code for GRMHD around rotating BHs and described by
\cite[Gammie et al. (2003)]{gammie03}. We use the values of
 $M$=3.4 $\times10^8$M$_\odot$ (estimated by Wilman et al. 2005)
and $a=0.93$ (arbitrarily chosen) 
for the mass and dimensionless spin parameter of the central supermassive BH. 
The accretion of the simulation is triggered by magneto-rotational instability in a hydrostatic equilibrium state, with the initial beta parameter $\beta =100$ and the specific heat ratio $\gamma=5/3$. The simulation is performed up to a
boundary radius of
$R_{max}=40 R_g$ with a computational resolution of 256$\times$256,
being $R_g=GM/c^2$ the gravitational radius.
We choose the snapshot at the time  $t=2010 R_g/c$ 
as the background environment for the CRs emission simulations. 

\begin{figure}[h]
\begin{center}
 \includegraphics[width=5in]{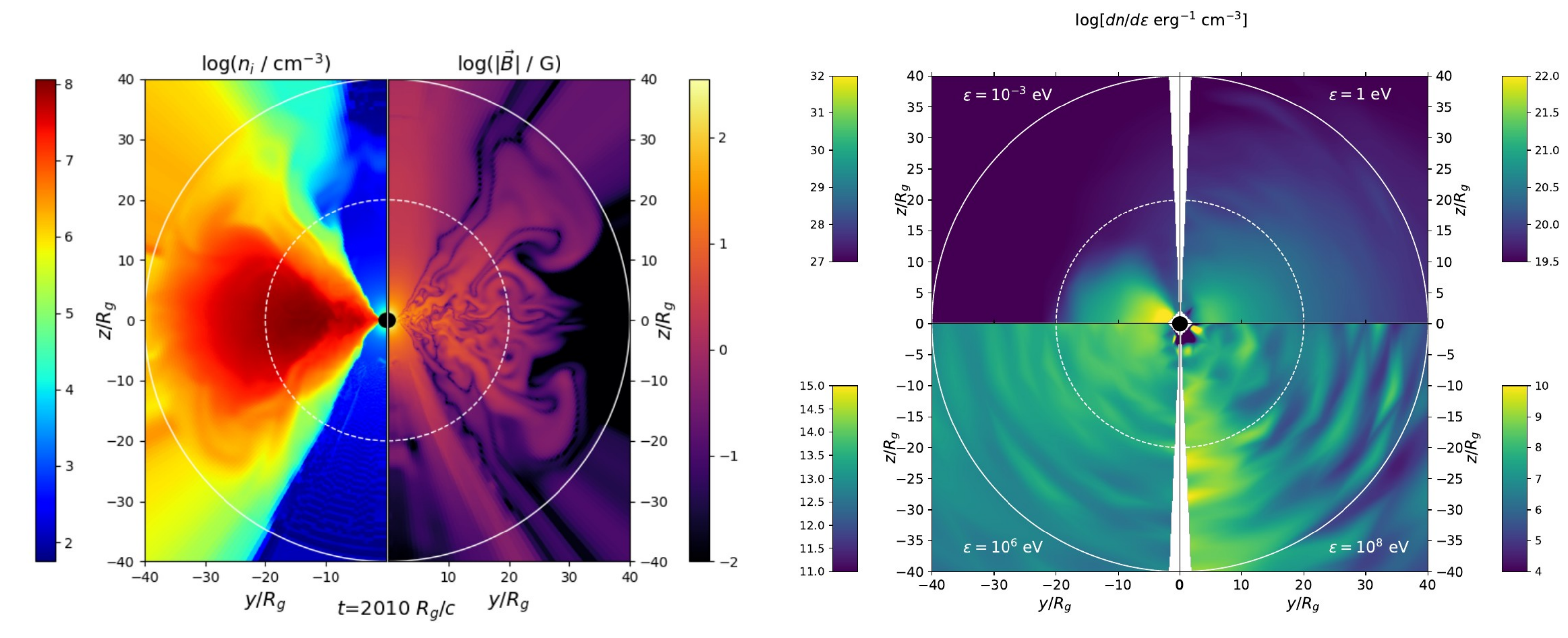} 
 \caption{
Left: Numerical GRMHD ADAF model for the core region of NGC1275. The figure shows  the gas number density 
and magnetic field profiles for the snapshot, obtained with the axi-symmetric {\tt harm} code.
Right: Photon field map of the snapshot flaw in the left panel, calculated with synchrotron+IC
radiative transfer, employing the {\tt grmonty} code. 
The inner white circle represents the
sphere inside which we inject the initial CRs accelerated by magnetic reconnection. 
The outer circle represents the spherical boundary where we detect neutrinos and $\gamma$-rays 
that result from the CR interactions plus electromagnetic cascades (see Section 3).
}
\label{fig1}
\end{center}
\end{figure}

The background photon field of the chosen snapshot flaw
is calculated performing post-processing synchrotron+inverse Compton (IC)
radiation transfer assuming the gas in a stationary state.
For this calculation we employ the publicly available Monte Carlo {\tt grmonty} code
(Dolence et al. 2009), 
using the proton-to-electron temperature ratio $T_p/T_e=3$.
For the mass unit we use the value of $M_{u}=3\times 10^{25} $g that we obtain by fitting
the calculated SED, corresponding to a line of sight with $\theta$=\ang{20} with respect to
the direction of the BH spin, to the X-ray component of the observed SED. 

\begin{figure}[b]
\begin{center}
 \includegraphics[width=5.3in]{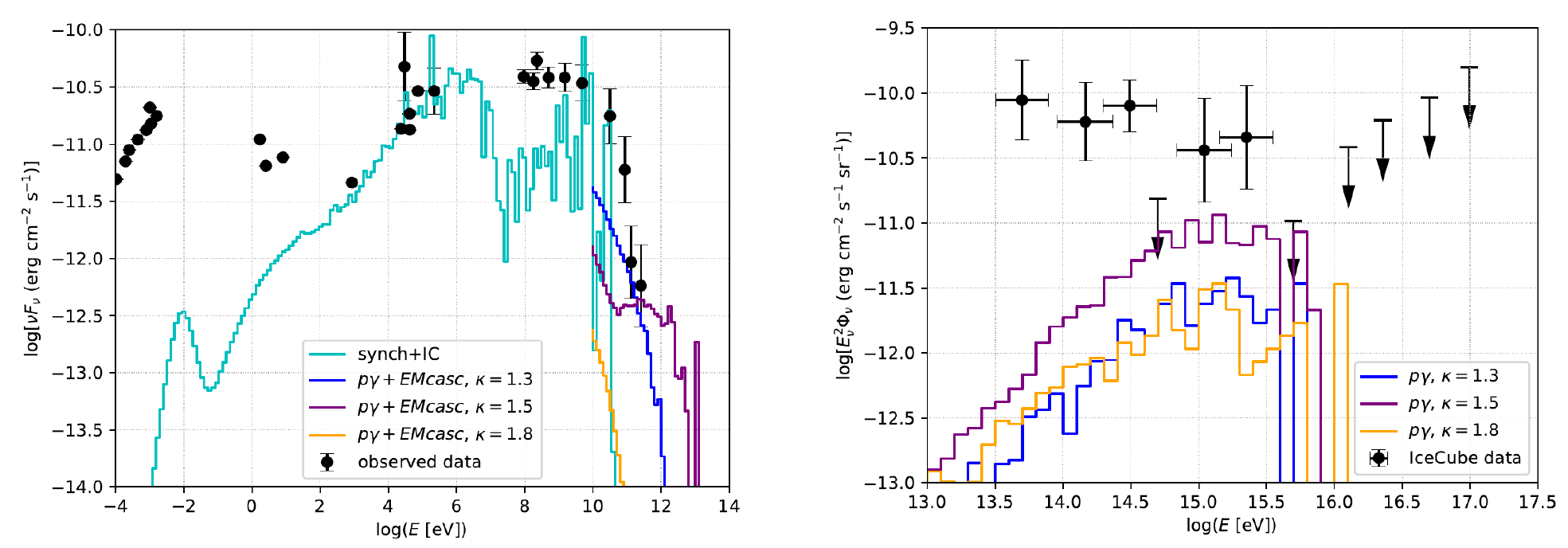} 
 \caption{
Left: SED for NGC1275, with leptonic synchrotron+inverse IC (cyan histogram, employing the {\tt grmonty} code) and with photo-pion processes plus electromagnetic cascading (blue, purple and orange histograms,
employing the {\tt CRPropa3} code).
The different histograms of the hadronic models correspond to the power-law index of the injected CRs.
The black points are observed data, from 
$MOJAVE$ (Lister et al. 2009),
$HST$ $FOS$ (Johnstone \& Fabian 1995),
$Swift$-BAT (Ajello et al. 2009),
$Fermi$-LAT (Abdo et al. 2009, Ackermann et al. 2012)
and
$MAGIC$ (Aleksi\'c et al. 2012a,b).
Right: Neutrino fluxes associated to the 
hadronic $\gamma$-rays emission models. The IceCube data (Aartsen et al. 2014) is shown for comparison.
}
\label{fig2}
\end{center}
\end{figure}

The gas density and magnetic field profiles of the chosen snapshot are shown in the left panel 
of Fig.\,\ref{fig1}, and its photon field map due to synchrotron + IC radiation are shown in the right panel.
The calculated SED of this leptonic emission model is displayed in the left panel of Fig.\,\ref{fig2} and
reproduces the observed data from 
X-rays to $\sim 10$ GeV $\gamma$-rays.

\section{Emission due to Interactions of CRs with the GRMHD Plasma Environment}

To calculate the neutrino and $\gamma$-ray emission due to CRs
in the accreting plasma profile obtained in the
last section, we employ the publicly available
{\tt CRPropa3} code (Alves Batista et al. 2016).
We consider the following interactions:
(i) photo-pion process ($p\gamma\rightarrow 2\gamma$ and $p\gamma\rightarrow ...+\nu_s$) 
that produces $\gamma$-rays and neutrinos, (ii) $\gamma$-ray absorption due to pair creation
($\gamma\gamma\rightarrow$ e$\pm$) that generates secondary electron/positron pairs, and 
(iii) IC scattering of background photons by the secondary electron/positron pairs.
Neutrinos produced due to photo-pion processes 
will escape the accreting torus travelling virtually unimpeded .

To emulate the release of CRs due to a magnetic reconnection event, we run simulations injecting
power-law distributions of CRs with the indices $\kappa=$1.3, 1.5 and 1.8, within the energy range
[100 TeV $-$ 50 PeV].
We consider the injection of protons within a sphere 
of $20 R_g$ (see Fig.\,\ref{fig1}).  
Particles stop being tracked when  
they complete a maximum trajectory length of $150 R_g$, or attain a minimum energy
of 10 GeV, or cross the detection sphere at $40 R_g$.
We calculate the neutrino and $\gamma$-rays fluxes that would be
detected at the Earth as $E_{\nu_s}^2\Phi_{\nu_s} =\frac{w}{4\pi}
\frac{\epsilon N_{\epsilon,\nu_s}}{\Delta\ln\epsilon}(4\pi R_0^2\Delta t)^{-1}$
and $\nu F_\nu = w\frac{\epsilon N_{\epsilon,\gamma}}{\Delta\ln\epsilon}
(4\pi R_0^2\Delta t)^{-1}$,
being $w=N_p/N^{sim}_p$ the ratio of physical to simulated CRs, $N_{\epsilon,\nu}$ and 
$N_{\epsilon,\gamma}$ the number of neutrinos and $\gamma$-rays
arriving at the detection sphere within the $\epsilon$ energy bin, 
$\Delta t$ the detection time interval, and $R_0=75$ Mpc the distance from NGC1275 to us.

The number of physical CRs, $N_p$, is expected to be a very small fraction of the number of thermal
ions $N_{th}$ in the background CRs injection region.
Thus, for the normalisation factor $w$ we use the upper limit condition $w \leq f_{CR} N_{th}/N_p^{sim}$
with $f_{CR}=10^{-5}$\footnote{We note that this factor of CRs is somewhat overestimated with regard to 
the value one obtains when considering the total magnetic reconnection acceleration power available in the 
simulated region (see e.g., Singh, de Gouveia Dal Pino \& Kadowaki, 2015). Nevertheless, for stronger background magnetic fields, 
we expect to obtain more realistic results, which will be explored in forthcoming work.}.
The number of thermal ions inside the injection sphere is estimated as
$N_{th}\sim(1/2)(4\pi/3)R_{inj}^3 \langle n_i \rangle$ (the 1/2 factor considers that most of the background ions
lie in half of volume of the injection sphere, see the gas density map in Fig.\,\ref{fig1}). 
Using $\langle n_i\rangle= 5\times 10^7$, and $N^{sim}_p=10^3$, 
we obtain the condition $w\leq 5\times10^{44}$.

In Fig.\,\ref{fig2} we show the $\gamma$-rays and  the associated neutrino fluxes 
 that result from the {\tt CRPropa3 } simulations and using $w=5\times 10^{44}$. 
The hadronic $\gamma$-ray fluxes models are the curves in the left panel that begin
at $10^{10}$ eV and correspond to different power-law indices of CR injection.
These hadronic emission fluxes tend to reproduce better the high energy tail of 
the SED for smaller power-law index.
The models of neutrino fluxes shown in the right panel correspond to the $\gamma$-ray curves
of the same colour.
The three models produce neutrinos peaking at $\sim$ PeV energies, producing fluxes that are
compatible with IceCube data (Aartsen et al. 2014).

\section{Summary and Conclusions}

We have investigated neutrino and $\gamma$-ray emission signatures 
produced in the core region of NGC1275, adopting a numerical GRMHD ADAF model for the
BH accreting plasma, together with  Monte Carlo simulations of leptonic and hadronic emissions.
The essential assumption for the hadronic component is the acceleration of 
CRs in a Fermi process by turbulent magnetic reconnection events (refs. in Section 1 and de Gouveia 
Dal Pino et al., these procs.), that we have emulated with the injection of power-law 
energy distributions of CRs. 

We have presented a model of pure leptonic synchrotron+IC emission that 
reproduces the observed SED in the
$10^3$-$10^{10.5}$ eV energy range  (see Fig.\,\ref{fig2}, left panel) which 
we have used as the target photon field for simulations of hadronic interactions. 
For the hadronic simulations we have
considered only photon-pion processes, from which we have 
obtained neutrino fluxes and  their associated emission due to $\gamma$-ray cascades
(see Figs. 1 and 2). 

The very high energy emission tail of NGC1275 is better reproduced 
with the hadronic models with smaller power-law index of CR injection,  
(which are compatible with the predictions of reconnection acceleration).
The associated  neutrino fluxes are consistent with the upper limits from IceCube 
(see Fig.\,\ref{fig2}, right panel) and from the models presented here, we conclude
that the core of NGC1275 would mostly contribute to neutrino detections at $\sim$PeV energies.
However, the neutrino fluxes presented in Fig.\,\ref{fig2} may be over-estimations, since they
come from a single source.

Possibly, other background plasma configurations as well as the inclusion of other hadronic interactions
(e.g. $pp$ and Bethe-Heitler processes) will produce better emission models. 
We will explore these possibilities in a forthcoming paper.
 
\section{Acknowledgments}

We acknowledge support from the Brazilian agencies FAPESP (2013/10559-5 grant) and CNPq (306598/2009-4 grant). 
The simulations presented in this lecture have made use of the computing facilities of the GAPAE group (IAG-USP) and the Laboratory of Astroinformatics IAG/USP, NAT/Unicsul (FAPESP grant 2009/54006-4). RAB is supported by the FAPESP grant 2017/12828-4 
and JCRR by the FAPESP grant 2017/12188-5.

\begin{thebibliography}{}

\bibitem[Aartsen, Ackermann, Adams. et al. (2016)]{Aartsen_etal16}
{Aartsen, M., Ackermann, M., Adams, J. et al.} 2016,
\textit{PhRvL} 113, 101101 


\bibitem[Abdo et al. \etal (2009)]{Abdo_etal09}
{Abdo, A. et al.} 2009,
\textit{ApJ}  699, 31


\bibitem[Ackermann et al. \etal (2012)]{Ackermann_etal12}
{Ackermann, M. et al.} 2012,
\textit{ApJS}  203, 4

\bibitem[Ajello et al. \etal (2009)]{Ajello_etal09}
{Ajello, M. et al.} 2009,
\textit{ApJ} 690, 367

\bibitem[Aleksi\'c \etal\ (2012)]{Aleksic_etal12}
{Aleksic, J. et al.} 2012,
\textit{A\&A}, 539L, 4 

\bibitem[Aleksi\'c \etal\ (2012)]{Aleksic_etal12}
{Aleksic, J. et al.} 2012,
\textit{A\&A}, 541, 99 



\bibitem[Alves Batista \etal\ (2016)]{RAB_etal16}
{Alves Batista, R., Dundovic, A., Erdmann, M., Kampert, K., Kuempel, D., Müller, G., Sigl, G., van Vliet, A., Walz, D. \& Winchen, T} 2016,
\textit{JCAP} 05, 38 

\bibitem[de Gouveia Dal Pino \etal (2018)]{gouveia18}
{de Gouveia Dal Pino, E. M. et al. } 
\textit{These Proceedings} 

\bibitem[de Gouveia Dal Pino \& Lazarian (2005)]{gouveia05}
{de Gouveia Dal Pino, E. M. \& Lazarian, A. } 2005,
\textit{A\&A}, 441, 845 

\bibitem[de Gouveia Dal Pino, Piovezan \& Kadowaki (2010a)]{gouveia10}
{de Gouveia Dal Pino, E. M., Piovezan, P. P. \& Kadowaki, L. H. S.} 2010a
\textit{A\&A}, 518, A5 

\bibitem[Dolence et al. 2009)]{gouveia10}
{Dolence, J., Gammie, C., Mościbrodzka \& M, Leung, P-K} 2009
\textit{ApJS}, 184, 387D 


\bibitem[Gammie et al. (2005)]{gammie05}
{Gammie, C., McKinney, J., Toth, G.} 2003, 
\textit{ApJ}, 598, 444

\bibitem[Johnstone \& Fabian (1995)]{Johnston95}
{Johnstone, R. M. \& Fabian, A. C.} 1995, 
\textit{MNRAS}, 273, 625

\bibitem[Kadowaki, de Gouveia Dal Pino, \& Singh (2015)]{kadowaki15}
{Kadowaki, L. H. S., de Gouveia Dal Pino, E. M. \& Singh C. B.} 2015, 
\textit{ApJ}, 802,113


\bibitem[Kadowaki, de Gouveia Dal Pino \& Stone (2018)]{kadowaki10}
{Kadowaki, L. H. S., de Gouveia Dal Pino, E. M. \& Stone J.} 2018a, 
\textit{arXiv:1803.08557}

\bibitem[Kadowaki, de Gouveia Dal Pino \& Stone (2018a)]{kadowaki10}
{Kadowaki, L. H. S., de Gouveia Dal Pino, E. M. \& Stone J.} 2018b, 
\textit{In prep.}


\bibitem[Kowal et al. 2011)]{Kowal_etal11}
{Kowal, G., de Gouveia Dal Pino, E. M. \& Lazarian, A.} 2011, 
\textit{ApJ}, 735, 102

\bibitem[Kowal et al. 2012)]{Kowal_etal12}
{Kowal, G., de Gouveia Dal Pino \& E. M., Lazarian, A.} 2012, 
\textit{PhRvL}, 108, 24

\bibitem[Lister, Aller, Aller et al. 2009)]{lister_etal09}
{Lister, M. L., Aller, H. D., Aller, M. F. et al.} 2009, 
\textit{AJ}, 137, 3718

\bibitem[Wilman et al. (2005)]{wilman05}
{Wilman, R., Edge, A., Johnstone, R.} 2005, 
\textit{MNRAS}, 359, 755

\end{thebibliography}
\end{document}